**The Clinical Trial Pipeline Reveals the Next Wave of Artificial Intelligence in Healthcare: A Multidimensional Analysis of 8,532 Registered Studies**

Lior Rokach

[1]The Stein Faculty of Computer and Information Science, Ben-Gurion University of the Negev, Israel

## Summary

Background: The prospective clinical evaluation of artificial intelligence in medicine has expanded rapidly, yet no comprehensive multidimensional analysis of the global AI clinical trial landscape exists. Understanding which AI technologies are being tested, in which clinical contexts, at what level of autonomy, and with what degree of translational maturity is essential for identifying where the evidence base is robust, where it is critically sparse, and where future investment should be directed.

Methods: We systematically identified all AI-related clinical trials registered in ClinicalTrials.gov using a broad keyword-based search followed by an LLM-based classifier. Utilizing few-shot learning, each identified trial was then classified across seven complementary analytical dimensions: clinical function, data modality, clinical specialty, degree of AI integration and autonomy, clinical workflow position, translational maturity stage, and AI epistemic role. A novel four-level autonomy taxonomy was applied to characterize the degree to which AI systems operate independently of human oversight, and a tripartite epistemic role framework was introduced to distinguish AI systems that are active clinical agents from those that serve as post-hoc analytical instruments or upstream design oracles.

Results: A total of 8,532 AI clinical trials were identified, spanning 32 clinical specialties and representing a 3.6-fold acceleration in registrations from 2019 onward, with 80% of all trials registered within this period. Of these, 30.5% employed a randomized controlled design. Imaging-based AI constituted the largest single modality (2,475 trials, 29%), while clinical text and NLP-based trials grew 7-fold between 2018 and 2025, the steepest trajectory of any modality and one coinciding temporally with the public availability of large language model platforms. Prognostic AI (4,324 trials) narrowly exceeded diagnostic AI (3,828 trials), reflecting a field-wide shift from disease detection toward risk stratification and trajectory prediction, while treatment recommendation AI remained the least developed task category (768 trials, 9%). Translational maturity analysis revealed that 3,259 trials (38%) remain in retrospective validation and 1,802 (21%) in silent prospective evaluation, indicating that the majority of the pipeline is still generating algorithmic rather than clinical evidence. A total of 184 trials were classified as Level 4 semi-autonomous or closed-loop AI systems, of which 68% were concentrated within closed-loop glucose management, with the remainder distributed across closed-loop anesthesia, respiratory management, vasopressor control, and neurostimulation. Multimodal AI systems — integrating two or more data streams — accounted for 33.6% of all trials, with imaging--omics (324 co-occurrences), imaging--physiological signals (445), and physiological signals--wearables (309) constituting the dominant fusion architectures. Geographic analysis revealed that West Europe, North America, and East Asia together account for 86% of all geolocated trials, while Africa, South America, and South Asia combined contribute under 5%. Specialty analysis identified severe underrepresentation of rheumatology (1.2% of trials), hematology (1.7%), infectious diseases (1.5%), and medical genetics (0.4%) relative to their global disease burden. The median trial enrollment was 193 participants, with 35% of trials enrolling fewer than 100 participants, and 83% of trials were sponsored by academic or non-profit institutions rather than industry.

<u>Conclusions:</u> Clinical AI has completed its first major transition — from retrospective algorithmic development to large-scale prospective evaluation — but has not yet completed the more consequential transition toward adequately powered, geographically representative, long-term outcome trials that can support guideline-level recommendations. Structural gaps in specialty coverage, geographic representation, trial scale, and translational maturity define the priorities for the next phase of clinical AI research. The near-exclusive concentration of semi-autonomous closed-loop AI in glucose management, the rapid emergence of language-based multimodal systems, and the growing separation between AI-as-clinical-agent and AI-as-analytical-instrument are the defining transitions visible in the prospective trial pipeline. Addressing these structural imbalances will determine whether the extraordinary breadth of ongoing AI clinical investigation translates into equitable and demonstrable benefit for patients.

## Introduction

Artificial intelligence (AI) in medicine is often discussed through benchmark performance, retrospective validation studies, and high-profile demonstrations. Yet, to anticipate what clinicians are likely to encounter in practice over the next decade, the answer may lie in the clinical trial ecosystem rather than in papers or patents.

Most clinical AI systems follow a recognizable translational pathway beginning with patents, proof-of-concept systems, or retrospective studies on curated datasets. These studies are often accompanied by reports of high accuracy, sensitivity, specificity, or area under the curve (AUC). Yet retrospective performance alone rarely changes clinical practice. Before large-scale deployment, AI systems must demonstrate robustness, workflow compatibility, safety, and clinical utility under prospective conditions before becoming part of routine care. Clinical trials, therefore, represent the critical transition from algorithmic promise to operational medicine[1]. As such, ongoing and planned trials provide a forward-looking view of the future clinical AI landscape. Moreover, clinical trials reflect where hospitals, regulators, industry, and healthcare systems believe AI is most likely to deliver meaningful clinical impact, and are therefore willing to invest substantial financial, operational, and regulatory resources to evaluate it prospectively under real-world conditions.

However, evaluating clinical AI prospectively is inherently more complex than evaluating conventional medical interventions. Unlike drugs with fixed molecular compositions, clinical AI systems are dynamic sociotechnical entities that may be continuously monitored, recalibrated, retrained, and updated in response to distributional shifts, vendor releases, user feedback, and evolving workflows[2]. As a result, prospective AI trials often evaluate moving targets rather than static interventions.

We sought to characterize the emerging clinical AI landscape through the lens of clinical trials. To this end, we identified all AI-related trials registered on ClinicalTrials.gov and classified them across multiple complementary dimensions capturing their technological, clinical, and translational properties. The analysis was designed to uncover the structural patterns underlying the AI trial pipeline: which domains attract the largest prospective investment, which data modalities dominate adoption, how autonomously AI systems are deployed relative to clinicians, and where major gaps between technological ambition and clinical evidence remain. To our knowledge, this represents the largest multidimensional analysis of AI-related clinical trials to date, encompassing 8,532 trials spanning more than two decades of prospective AI evaluation in medicine.

Specifically, the analysis pursued the following objectives:

1. To identify AI-related clinical trials registered in ClinicalTrials.gov using a systematic keyword-based strategy and characterize their distribution across time, geography, sponsor type, and study design.
2. To classify each trial across seven complementary dimensions: clinical function, data modality, clinical specialty, autonomy level, workflow position, translational maturity, and AI epistemic role.
3. To quantify temporal trends in trial registration across domains, modalities, and maturity stages in order to identify emerging frontiers, saturated niches, and underserved areas.
4. To characterize translational maturity by distinguishing trials focused on algorithmic performance from those evaluating clinical outcomes, and to identify where the pipeline is most concentrated and constrained.
5. To identify structural gaps in the AI clinical trial ecosystem, including underrepresented specialties, geographically concentrated evidence bases, and the near-absence of prospective evaluation for semi-autonomous and generative AI systems.

## Methods

The overall study design and analytical workflow are illustrated in Figure 1. Briefly, all registered studies from ClinicalTrials.gov were first screened using an inclusive AI-related keyword strategy, followed by a few-shot learning refinement stage to identify genuinely AI-related trials. Structured metadata and free-text descriptions were subsequently used to extract study characteristics and classify each trial across seven orthogonal dimensions capturing clinical function, modality, specialty, workflow position, autonomy level, translational maturity, and epistemic role.

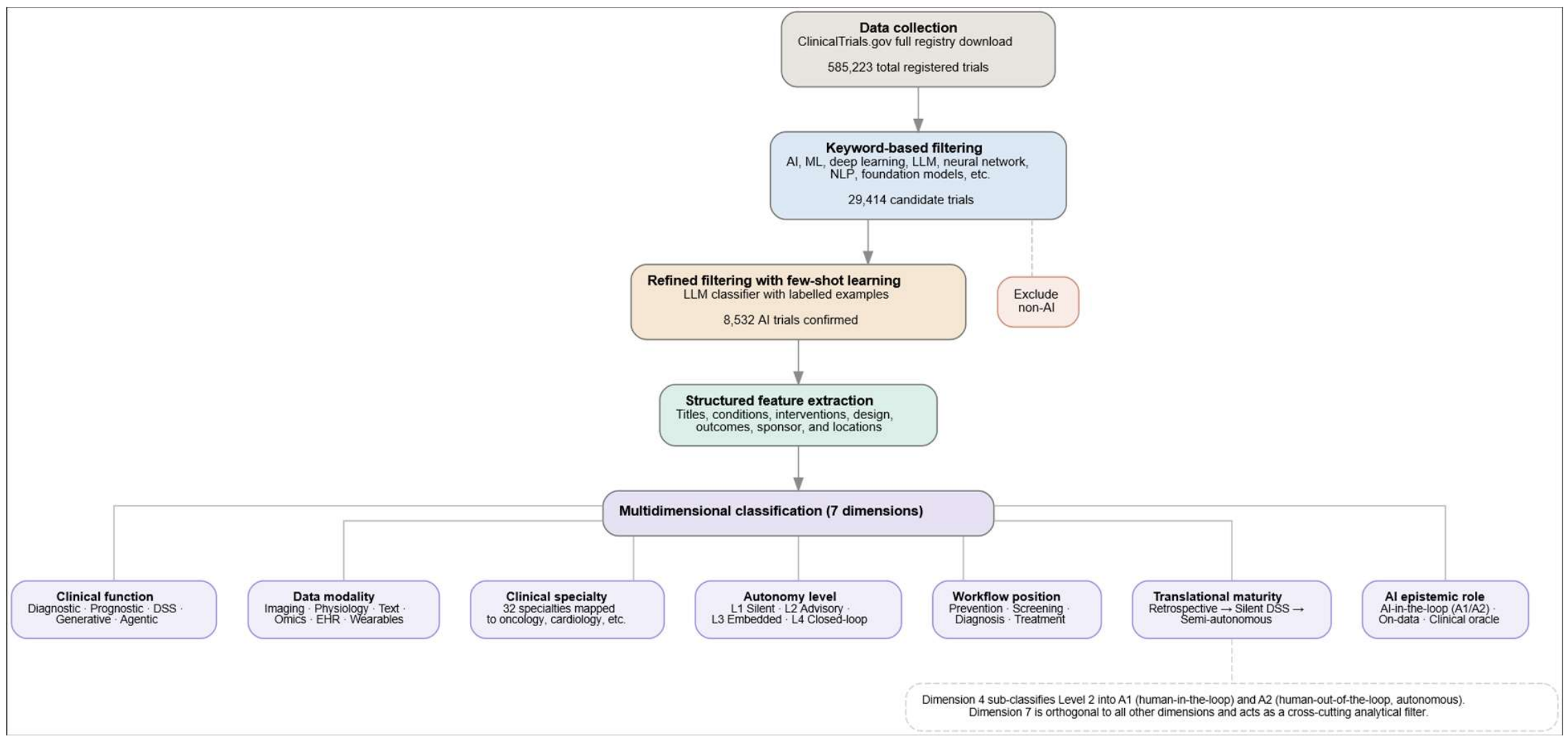

Figure 1: Overview of the study workflow for identification and multidimensional characterization of AI-related clinical trials from ClinicalTrials.gov

Filtering process

First, we filtered all clinical trials in ClinicalTrials.gov using an inclusive list of keywords (see Supplementary 1). The list combined highly specific AI-related terms, such as “convolutional neural network,” which offer high precision, with broader terms such as “model” and “algorithm,” which are not uniquely associated with AI. We then use few-shot learning with Claude Sonnet 4.6 to precisely identify all AI-related trials. Trials assigned an AI-related probability of at least 95% Claude's classification was classified as positive. Trials assigned a probability of 20% or lower were classified as negative. All remaining cases were treated as ambiguous and were reviewed using a second LLM (Gemini 3.1 Pro preview). Trials classified as AI-related with high confidence (> 80%) by both models were retained. Discrepant or low-confidence cases underwent manual review before final inclusion.

Ambiguity in identifying AI-related trials is further complicated by the fact that AI is both a transformative technology and a continuously evolving concept. Its definition has shifted repeatedly over the years: from symbolic AI to machine learning to generative AI, and, more recently, discussions surrounding agentic AI. For example, MYCIN is widely regarded as one of the first major attempts to apply AI in healthcare. Developed in the 1970s, MYCIN relied on manually encoded rules and expert knowledge rather than statistical learning, yet at the time it represented the cutting edge of “artificial intelligence.” Decades later, many systems that would once have been labeled AI became so routine that they were reclassified simply as software or advanced analytics.

It should be noted that trials evaluating drugs originally discovered or designed using AI were excluded from the AI-related trial cohort when the corresponding clinical trial registry did not explicitly reference AI methods. This reflects a broader limitation of registry-based analyses, where the technological origin of a therapeutic candidate may become invisible once the intervention enters conventional clinical development pipelines.
For example, INS018_055 (Rentosertib), a therapy for idiopathic pulmonary fibrosis discovered by Insilico Medicine using AI-driven drug discovery models, is currently being evaluated in clinical trials such as NCT05154240 (Registration Date: 11/17/2021), yet the associated registry entries do not mention artificial intelligence, machine learning, or related computational methodologies.
Consequently, such studies could not be systematically identified through registry natural language processing despite representing some of the most advanced real-world translational successes of AI in biomedicine.

A Taxonomy for the Multi-Dimensional Characterization of AI Clinical Trials

To enable a systematic and multidimensional characterization of the clinical AI trial landscape, each registered trial was assigned values across a structured set of analytical dimensions. Rather than treating the corpus as a flat enumeration of studies, this framework was designed to capture the full complexity of how AI systems vary in their clinical purpose, the data they consume, the medical domains they target, the degree of autonomy they exercise, and the stage of translational evidence they represent. Some dimensions are primary and substantive, i.e. they define what kind of AI system is being evaluated and where in the care pathway it operates. Others are secondary and methodological, i.e. they characterize the rigor of the evidence being generated and the conditions under which that evidence can be expected to generalize. A final set of cross-cutting dimensions serves as stratifiers applicable across all of the above, enabling detection of concentration effects by geography, sponsor type, and temporal trend. Together, these dimensions form a coordinate system within which any individual AI clinical trial can be precisely located, and through which aggregate patterns, dominant modalities, neglected specialties, maturity bottlenecks, and emerging frontiers, become analytically visible. The following sections describe each dimension in turn, with full coding specifications provided in Supplementary 2.

**Clinical Function.** Each trial was classified according to its primary AI function within the clinical encounter, spanning six categories: diagnostic AI, predictive and prognostic modeling, clinical decision support, generative and language-based AI, agentic or semi-autonomous systems, and operational or administrative AI. Diagnostic systems encompassed technologies directed at disease detection, classification, staging, or screening. Predictive systems included models estimating future clinical outcomes such as mortality, complication risk, treatment response, or disease trajectory. Clinical decision support systems captured AI tools assisting with therapeutic selection, triage, workflow prioritization, or management optimization. Generative AI covered large language model-based systems designed for clinical documentation, summarization, patient communication, or conversational interaction. Agentic AI referred to systems capable of semi-autonomous operation, adaptive intervention, or closed-loop physiological control. Operational AI included technologies designed to optimize healthcare logistics, scheduling, staffing, or resource allocation.

**Data Modality.** Trials were further characterized according to the primary data modality processed by the AI system. Eleven modality classes were defined: medical imaging, physiological signals, laboratory and biochemical data, omics data, clinical text and natural language, structured electronic health record data, wearable and remote monitoring data, behavioral and cognitive data, patient-reported outcomes, administrative and claims data, and surgical and procedural data. Obviously, this dimension also identifies multimodal systems that involve at least two modalities.

**Clinical Specialty.** Each trial was mapped to one or more primary clinical specialties based on the conditions, interventions, and keywords fields of the registry record. Seventeen specialty domains were defined, including cardiology, oncology, hematology, neurology, gastroenterology, radiology, surgery, critical care, emergency medicine, endocrinology, pulmonology, infectious diseases, psychiatry, ophthalmology, dermatology, rehabilitation medicine, and primary care. This dimension aims to identify specialties that are underrepresented relative to their disease burden and, thus, define priority areas for future investment.

**Integration and Autonomy Level.** A four-level ordinal taxonomy was applied to characterize the degree of clinical integration and the autonomy of the AI system within each trial. Level 1 systems operated in a silent or purely observational mode, without influencing clinical care. Level 2 systems provided advisory recommendations while clinicians retained full decision-making authority. Level 3 systems were embedded directly within clinical workflows through automated prioritization, alerts, or workflow orchestration. Level 4 systems constituted semi-autonomous or closed-loop technologies capable of dynamically modifying treatment or intervention parameters without mandatory per-decision human review. Within Level 2, a further distinction was drawn between AI-as-decision-support configurations (A1), in which a human clinician adjudicates every AI recommendation, and AI-as-autonomous-actor configurations (A2), in which the AI output directly triggers a clinical action without mandatory human review.

**Clinical Workflow Position.** Trials were classified according to their position within the care pathway, spanning prevention, screening, diagnosis, treatment selection, procedural guidance, monitoring, rehabilitation, and longitudinal follow-up. This axis is conceptually distinct from clinical function: a single specialty may simultaneously host trials across all workflow positions, and a single trial may span multiple adjacent positions. Screening and diagnostic positioning together constitute the majority of current trials, but a measurable and growing proportion address downstream positions --- treatment optimization, intraoperative guidance, and continuous post-discharge monitoring --- reflecting a broadening of AI's intended role from episodic classification toward longitudinal care integration.

**Translational Maturity Stage.** Each trial was evaluated according to its position within a five-stage translational maturity ladder: retrospective validation, silent prospective evaluation, interventional decision-support trial, workflow-integrated deployment, and semi-autonomous clinical operation. This staging framework aims to capture the progressive increase in evidential rigor and clinical consequence required to move from algorithm development to routine care integration.

**AI Epistemic Role.** This dimension was introduced to characterize where in the translational pipeline the AI system exerts its influence relative to the trial itself. Three epistemic roles were defined. Level A (AI-in-the-loop) designates trials in which the AI system is the active intervention or a constitutive component of it, with model outputs directly entering the clinical process in real time; this level was further decomposed into A1 (human-in-the-loop, advisory) and A2 (human-out-of-the-loop, autonomous). Level B (AI-on-the-data) designates trials in which AI plays no role in the intervention but is applied post hoc or in parallel to extract patterns, biomarkers, or predictive signals from trial data that conventional methods would not surface. Level C (AI-out-of-the-loop) designates trials in which AI contributed to the genesis of the intervention --- its discovery, optimization, or rational design --- but is no longer operationally present during the trial itself.

**Sponsor Type and Geography.** Trial sponsor class and geographic origin were extracted as cross-cutting stratifiers applicable across all analytical axes. Sponsor type was coded as academic/institutional, industry, government, or other, based on the lead sponsor classification field. Geographic origin was assigned from the locations module of each registry record.

Features Extraction and Taxonomy Mapping

Certain features (such as start date or recruitment status) were extracted directly from the JSON responses returned by the ClinicalTrials.gov API. However, most of the dimensions described above were not available as structured fields and were therefore inferred from the free-text trial descriptions, intervention details, study objectives, and eligibility criteria. In particular, medical specialties, AI data modalities, clinical task types, workflow positions, and levels of clinical integration were derived using few-shot learning with Anthropic's Claude, guided by a predefined taxonomy based on the trial narrative and intervention characteristics. Similarly, AI-related trials were further categorized according to their epistemic role, autonomy level, and translational maturity stage based on the described function of the AI system within the clinical workflow When structured and textual fields were incomplete or inconsistent, supplementary contextual information from sponsor descriptions, study summaries, and location metadata was used to resolve ambiguities.

## Results

As of the end of April 2026, we have identified 8,532 AI-related clinical trials. As expected, the single most striking finding is the rate of acceleration, showing that AI is moving from exploratory experimentation into a decisive evidential phase. Seventy-eight percent of all registered AI trials (6,635 out of 8,532) were initiated from 2020 onward, representing a 3.6-fold acceleration relative to the entire preceding period (2003–2019). The compound annual growth rate from 2015 to 2025 exceeds 20% per year. Crucially, this growth is not slowing: 2026 already shows 793 new registrations with the year still incomplete, making it almost certainly the highest single-year count in history.

The active recruitment pipeline provides an indication of where the field is heading. Of the 8,532 AI clinical trials identified, 3,450 (40.4%) constitute an active evidence-generation portfolio: 1,834 trials are currently enrolling patients, 969 are registered with confirmed start dates of which 480 are scheduled to begin recruitment in 2026 alone, 208 are enrolling by invitation, and 439 have completed enrollment but remain in active follow-up. Together, these studies carry a median planned enrollment of 250 participants and represent a projected aggregate patient exposure approaching 5.3 million individuals across the not-yet-recruiting cohort alone.

The current recruiting list includes a variety of AI-related trials. For example, one trial (NCT05437237) deploys a dry-electrode EEG headset in the ambulance, with an AI algorithm aiming to distinguish ischemic stroke from stroke mimics (i.e. a non-vascular medical condition that resembles seizures or hypoglycemia). It could redirect ambulances to comprehensive stroke centres rather than the nearest hospital. In another trial (NCT07362433), an AI-enabled digital stethoscope is being used in pediatric patients presenting with upper respiratory infections to support AI-guided antibiotic stewardship, with the primary goal of reducing unnecessary antibiotic prescriptions and, consequently, limiting the emergence of antimicrobial resistance.  A particularly ambitious trial (NCT07132151) employs a reinforcement learning algorithm for real-time autonomous control of anesthetic drug dosing delivery during a surgical abortion procedure without involvement of an anesthesiologist in the decision. The AI system continuously monitors vital signs and induction-depth signals, dynamically adjusting infusion rates to maintain the patient within a predefined sedation window from induction through recovery. In the control arm, the anesthetic drug regimen is administered and titrated by experienced anesthesiologists.

Prior to 2020, interventional studies consistently outnumbered observational ones. From 2020 onward, observational studies overtook interventional ones, peaking in 2022–2024 before the gap begins to narrow again in 2025. This is likely explained by two concurrent forces. First, the COVID-19 pandemic made prospective randomized recruitment operationally difficult, redirecting resources toward retrospective and silent-prospective designs. Second, the proliferation of AI tools embedded in EHR systems facilitated passive observational deployment at scale. However, 2025 shows a clear reversal: interventional trials (648) are now once again approaching parity with observational ones (583), suggesting a maturing pipeline in which institutions are regaining confidence in prospective evaluation. The overall split remains nearly equal: 50.8% interventional versus 49.2% observational.

Specialty landscape: who is investing in AI and who is not

Otolaryngology (ENT) leads with 1,994 trials - a finding likely driven by the large volume of endoscopy-guided AI systems (laryngoscopy, cochlear implant optimization, and audiological screening), which are highly amenable to standardized image-based evaluation. Oncology follows with 1,295 trials, reflecting decades of sustained investment in diagnostic imaging, biomarker discovery 9and treatment planning AI. Cardiology (750), neurology (637), endocrinology (579), and pulmonology (534) complete the high-activity tier.

Endocrinology shows the highest proportion of interventional decision-support trials (52%) and is the only specialty with three semi-autonomous (closed-loop) systems — all artificial pancreas devices for Type 1 diabetes. This reflects a well-established regulatory pathway (the FDA's De Novo authorization for automated insulin delivery), which other specialties have not yet achieved. Neurology and gastroenterology show relatively balanced distributions across maturity stages, suggesting these fields are progressing steadily through the translational pipeline. In contrast, oncology and pulmonology remain disproportionately retrospective (48% and 49%, respectively), indicating that, despite large absolute trial volumes, most of the activity remains in algorithmic development rather than prospective clinical evaluation.

Several high-burden specialties are strikingly underrepresented. Rheumatology accounts for only 1.2% of trials, despite the complexity and chronicity of diseases such as rheumatoid arthritis and systemic lupus erythematosus, for which AI-assisted monitoring could deliver measurable benefits. Hematology represents only 1.7%; Infectious diseases (1.5%), nephrology (2.2%), medical genetics (0.4%), and allergy and immunology (0.3%) are similarly underdeveloped. These omissions represent both a scientific gap and an investment opportunity.

Oncology leads the active pipeline (actively recruiting or about to recruit) in absolute volume (662 trials, 51% activation rate), with neurology (312, 49%), pulmonology (258, 48%), and cardiology (348, 46%) reflecting the highest proportional commitment to prospective evaluation relative to their historical trial base, while infectious diseases (29% activation) and endocrinology (30%) lag behind: the former through chronic underinvestment, the latter through the maturation of the artificial pancreas literature into completed rather than ongoing studies.

### Task type: the shift from diagnosis to prognosis

Prognostic AI narrowly leads diagnostic AI (4,324 vs. 3,828 trials), a finding that marks a qualitative shift in the field. Earlier generations of clinical AI were predominantly diagnostic. The current pipeline reflects a broader ambition: predicting future events, estimating trajectories, and stratifying risk. Patient monitoring tasks rank third (1,715 trials), consistent with the expansion of AI into continuous surveillance. Treatment recommendation trials were significantly higher (768 trials, 9%), underscoring that prescriptive AI (systems that actively recommend specific interventions) remains the least developed and arguably the most consequential frontier. Administrative automation (900 trials) is growing rapidly, driven by LLM-based documentation and workflow tools.

The task distribution of active trials reveals a prognostic shift more pronounced than in the full corpus: prognosis leads with 1,906 active trials versus 1,599 diagnostic trials, while treatment recommendation remains the least represented active task category at only 294 trials — confirming that prescriptive AI is the most underdeveloped frontier in both the historical and forward-looking pipeline.

### Data modalities: the patient-reported outcomes anomaly and the NLP surge

Among the analytically meaningful modalities, imaging leads at 2,475 trials (29%), confirming its role as the primary gateway for AI adoption (see Figure 2). Physiological signals rank second (1,641), driven largely by ECG-based and continuous monitoring systems. Clinical text and NLP rank fourth (1,219), but their growth trajectory is the steepest. From 36 trials in 2018, NLP-based trials rose to 252 by 2025 (a 7-fold increase), the fastest of any modality, and one that has clearly accelerated since 2022, coinciding with the public release of large language models. Wearable and remote monitoring trials (693) have also grown steadily, more than 7-fold over the same period. Omics-based trials (647) show sustained growth, reaching 99–106 per year in 2023–2024.

A trial was classified as multimodal when its AI system integrated two or more distinct data modalities drawn from a predefined set of nine core categories: imaging, physiological signals, laboratory and biochemical data, omics, clinical text and natural language, structured EHR data, wearable and remote monitoring data, behavioral and cognitive data, and surgical and procedural data. Patient-reported outcomes, administrative data, and claims data were excluded from the modality count for this analysis, as their near-universal co-occurrence with substantive modalities would systematically inflate multimodal prevalence estimates. Across the 8,532 AI trials analyzed, 2,867 (33.6%) integrated two or more core modalities, 954 (11.2%) integrated three or more, and a small but growing cohort of 57 trials (0.7%) integrated five or more distinct data streams within a single AI system. The proportion of trials classified as multimodal has risen measurably over the past five years, from 27.8% in 2020 to 36.4% in 2025. More informative than the proportion is the absolute volume: the annual count of newly registered multimodal trials increased from 79 in 2015 to 448 in 2025, a 5.7-fold increase that substantially outpaces the overall 5.6-fold growth in AI trial registrations over the same period, indicating that multimodal integration is growing at least as rapidly as AI trial activity in aggregate.

Among the 2,867 multimodal trials, analysis of pairwise modality co-occurrences reveals a structured set of dominant combinations that reflect the natural data availability and clinical logic of different trial contexts (Table 1). The most frequent pairing is imaging combined with physiological signals (445 co-occurrences), which characterizes a large class of cardiothoracic, neurological, and surgical trials in which real-time monitoring data complement static or dynamic imaging.

Table 1: Top multimodalities in AI-related clinical trials.

| **Modality 1** | **Modality 2** | **Co-occurrences** |
|---|---|---|
| Imaging | Physiological signals | 445 |
| Imaging | Surgical & procedural | 389 |
| Imaging | Omics | 324 |
| Physiological signals | Wearable & remote monitoring | 309 |
| Text / NLP | Behavioral & cognitive | 305 |

| Imaging | Laboratory | 305 |
|---|---|---|
| Physiological signals | Behavioral & cognitive | 302 |
| Imaging | Text / NLP | 276 |
| Imaging | Behavioral & cognitive | 269 |
| Physiological signals | Text / NLP | 245 |
| Physiological signals | Surgical & procedural | 222 |
| Physiological signals | Laboratory | 209 |
| Imaging | EHR structured data | 159 |
| Text / NLP | EHR structured data | 158 |

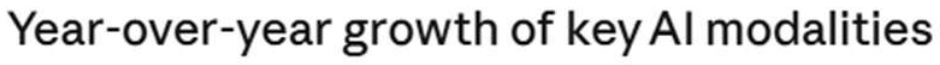

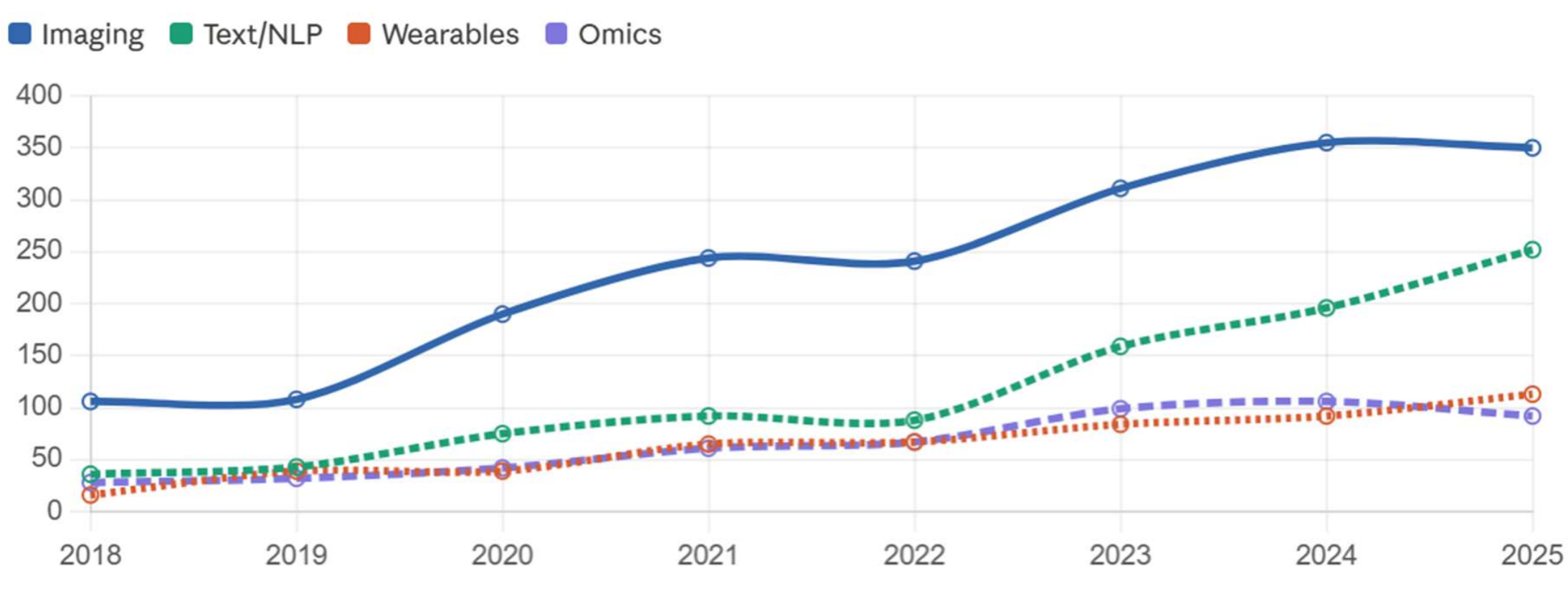

Figure 2: Emerging modality trends (2018–2025)

<u>Translational maturity: most of the pipeline is still pre-deployment</u>

The plurality of trials falls within the interventional decision-support category (n = 3,368), meaning that AI is evaluated as a workflow-embedded recommendation tool but have not yet demonstrated routine autonomous operation. A substantial fraction (3,259 trials) remains in retrospective validation, generating algorithmic rather than clinical evidence. Silent prospective evaluation accounts for a further 1,802 studies, in which AI systems observe clinical data in real time but do not yet influence care. Taken together, these three stages encompass the vast majority of the current pipeline, reflecting a field that is still accumulating the prospective evidence base required for large-scale deployment. At the frontier of the maturity spectrum, the dataset contains 184 trials classified as Level 4 semi-autonomous or closed-loop AI systems, mostly from the last 5 years. This reveals both the genuine progress and the persistent concentration of advanced AI autonomy within a narrow set of physiologically tractable domains. The overwhelming majority of these trials (125 trials) involve closed-loop glucose management in Type 1 and Type 2 diabetes through the artificial pancreas paradigm, which benefits from three decades of control-theoretic development, a well-defined physiological target variable, and an established FDA regulatory pathway for automated insulin delivery devices. A second cluster of 10 trials involves closed-loop anesthesia and sedation control guided by depth-of-anesthesia monitors such as the bispectral index, including closed-loop propofol delivery in pediatric cardiac surgery and closed-loop total intravenous anesthesia. A third group covers closed-loop respiratory management --- automated oxygen titration in preterm infants, closed-loop mechanical ventilation, and ECMO-integrated control. A fourth emerging cluster involves closed-loop vasopressor delivery for intraoperative hemodynamic stabilization, closed-loop antibiotic infusion, and brain-computer interface-driven neurostimulation for stroke rehabilitation.

Sitting at the outermost edge of this frontier is trial NCT07132151.
This randomized study evaluates a reinforcement learning algorithm designed for fully autonomous control of ciprofol dosing during surgical abortion, directly comparing the system against experienced anesthesiologists while removing physician involvement from moment-to-moment dosing decisions.
Unlike the model-predictive control systems that dominate much of the artificial pancreas literature, reinforcement learning relies on a non-deterministic and continuously adaptive policy that updates according to patient feedback, placing the system conceptually closer to a learning agent than to a conventional programmable medical device.

Across all 184 Level 4 trials, the striking scarcity of semi-autonomous AI systems outside endocrinology, anesthesiology, critical care, and neurostimulation remains one of the defining structural characteristics of the current landscape. In domains where a stable physiological signal can serve as a continuous feedback variable, such as glucose levels, anesthetic depth, oxygen saturation, or hemodynamic pressure, closed-loop AI systems have begun to establish their first clinical footholds. In nearly all other medical settings, however, the translational pipeline has not yet advanced to this degree of autonomy.

Geography: a three-region concentration with a global equity problem

Western Europe (33%), North America (27%), and East Asia, predominantly China (21%), account for 81% of all trials, while Africa, South Asia, and South America combined represent under 5% of the global AI clinical trial portfolio. This concentration has multiple implications. First, it raises fundamental questions about the generalizability of systems trained and validated predominantly in high-income, high-resource healthcare environments. Second, the global equity problem is stark: Africa accounts for 2.2% of trials despite bearing a disproportionate burden of infectious diseases, maternal mortality, and non-communicable diseases. South America represents 1.5% and South Asia (including India with over 1.4 billion people) only 1.0%. Third, China's growing share of East Asian trials (the majority of the 1,813 in that region) represents a significant scientific input, but the translational pathway from Chinese registry trials to international clinical adoption remains unclear.

## Discussion

This analysis of 8,532 AI-related clinical trials registered in ClinicalTrials.gov represents, to our knowledge, the largest multidimensional characterization of the prospective clinical AI pipeline assembled to date. The results show that clinical AI has moved well beyond retrospective algorithm development and is now generating prospective evidence at a scale that would have been difficult to imagine even a few years ago.
At the same time, the data expose major structural imbalances in specialty coverage, geographic representation, translational maturity, and evidential quality that must be addressed before the benefits of clinical AI can be broadly and equitably realized.

The 3.6-fold increase in trial registrations since 2019, with 78% of all trials initiated during this period, reflects more than a bibliometric trend.
It reflects the convergence of several enabling forces: mature deep learning architectures, large digitized clinical datasets, clearer regulatory pathways for Software as a Medical Device, and more recently, the emergence of large language models extending AI from structured pattern recognition into natural language understanding and generation.

The 2026 registration rate, already the highest on record despite the year being incomplete, suggests that this expansion has not yet plateaued. Importantly, interventional trials in 2025 are again approaching parity with observational studies, reversing the post-COVID dominance of observational designs. This pattern is consistent with a field transitioning from exploratory, retrospective research to prospective, hypothesis-driven evaluation.
Such a transition is critical because the true utility of AI systems only becomes visible under real-world conditions involving missing data, workflow interruptions, alert fatigue, and heterogeneous patient populations — conditions that retrospective studies rarely capture adequately.

Imaging-based AI remains the dominant modality in the trial corpus (2,475 trials, 29%), reflecting decades of computer vision research in radiology, pathology, ophthalmology, endoscopy, and surgical video analysis. However, the most important modality-level trend is the rapid expansion of text and NLP-based systems. Trials involving language AI increased 2.8-fold between 2018--2021 and 2022--2025 (246 versus 695 trials), coinciding almost exactly with the emergence of foundation models and large language models after 2022.
This likely underestimates the true pace of adoption, as many institutional deployments of language AI occur under quality improvement frameworks that do not require formal trial registration. Wearable monitoring and omics-based trials also nearly doubled during the same period, suggesting a broader transition from isolated single-modality classifiers toward longitudinal multimodal AI systems operating continuously across the patient journey.

The finding that prognostic AI trials (4,324) now slightly exceed diagnostic trials (3,828) reflects more than a numerical shift. Early clinical AI focused primarily on classification: given a signal, identify the disease category. The current pipeline increasingly focuses on predicting future patient trajectories. This transition toward risk stratification and personalized prediction is especially visible in oncology, cardiology, and hematology, where prognostic AI systems are beginning to enter prospective treatment decision workflows.
In contrast, treatment recommendation systems remain relatively uncommon (768 trials, 9%). This gap is particularly important because prescriptive AI (i.e. systems that actively recommend interventions, doses, or management strategies) likely represents the area of greatest potential clinical impact and weakest prospective evidence base. The regulatory, ethical, and liability frameworks needed to support such systems at scale remain substantially underdeveloped, and the trial landscape reflects this limitation.

The maturity distribution reveals a paradox. Most trials remain concentrated in retrospective validation (3,259 trials) or silent prospective evaluation (1,802 trials), indicating that much of the field is still generating algorithmic rather than clinical evidence. A model that performs well retrospectively or during silent deployment may demonstrate technical competence, but it has not yet shown that it changes clinician behavior, improves patient outcomes, or justifies deployment costs. At the same time, the identification of 184 Level 4 closed-loop and semi-autonomous trials demonstrates that parts of the pipeline have already reached substantial technical and regulatory maturity. The artificial pancreas ecosystem, accounting for approximately 130 of these trials over more than two decades, represents the clearest example of the translational arc from algorithm to fully deployed autonomous medical system.
Its progression from tightly monitored feasibility studies to FDA-cleared home-use systems provides both a roadmap and a warning regarding the timescales required for safe autonomous deployment. The appearance of NCT07132151, a reinforcement learning-based autonomous anesthesia system randomized directly against expert anesthesiologists, suggests that this paradigm may soon extend beyond glucose management into broader clinical domains.

The specialty distribution is markedly skewed and does not mirror global disease burden. ENT and otolaryngology (1,994 trials) and oncology (1,295 trials) together account for nearly 39% of all trials, largely because these specialties generate large, standardized imaging datasets that are ideally suited for deep learning. In contrast, rheumatology, hematology, infectious diseases, nephrology, and medical genetics collectively account for fewer than 7% of trials despite affecting hundreds of millions of patients worldwide. This imbalance reflects a broader structural reality: AI tends to follow the path of least data resistance. Fields with large digitized datasets and clearly defined pattern-recognition tasks attract rapid adoption, whereas specialties characterized by heterogeneous disease processes, sparse structured data, or poorly standardized outcomes remain underrepresented. Correcting this imbalance will require substantial investment in longitudinal registries, standardized phenotyping, and federated data-sharing infrastructure before meaningful algorithmic progress can occur.
The underrepresentation of hematology is particularly striking given the richness of its multi-omics, cytometry, and longitudinal monitoring data, as well as the potential of AI to integrate these signals into dynamic prognostic systems beyond the capabilities of current clinical scores.

Western Europe, North America, and East Asia account for 81% of all geolocated AI trials. This is not merely a geographic imbalance but a looming generalizability problem.
AI systems trained primarily in high-resource healthcare systems inevitably encode the characteristics of those systems, including patient demographics, imaging hardware, laboratory standards, coding practices, and patterns of care. When deployed in substantially different healthcare environments, performance degradation is common. Africa, South Asia, and South America collectively contribute fewer than 5% of AI trials despite representing a major share of the global burden of infectious disease, maternal mortality, and chronic disease. A global AI evidence base built predominantly on Western and East Asian populations risks systematically mischaracterizing model performance in populations that may benefit most from deployment. Addressing this gap will require targeted funding, international trial infrastructure, and regulatory incentives supporting validation in low- and middle-income settings.

Two structural limitations are particularly notable. First, 35% of trials enroll fewer than 100 participants, generally insufficient for demonstrating clinical benefit, subgroup robustness, or reliable safety estimates. The median enrollment of 193 participants suggests that the field remains dominated by feasibility and technical performance studies rather than adequately powered effectiveness trials. Second, the proportion of randomized controlled trials declined from approximately 33% in 2018--2019 to 26% in 2021--2023 before recovering toward 34% in 2025. Although part of this decline likely reflects post-COVID disruptions, it also highlights the field's broader tendency to favor observational evaluation over rigorous experimental designs. As a result, most current AI trials remain poorly positioned to answer the questions that matter most to clinicians, regulators, and payers: whether these systems improve outcomes, reduce costs, perform equitably across populations, and remain reliable over time as clinical environments evolve.

**Data availability:** The data can be downloaded from:
https://drive.google.com/file/d/1CnPjV9luHEgzsgROjsqwKGoPjpDw6FiI/view?usp=sharing

**Declaration of generative AI and AI-assisted technologies in the manuscript preparation process**

During the preparation of this work, the author used ChatGPT in order to proofread the paper and generate Figure 1. After using this tool/service, the author reviewed and edited the content as needed and takes full responsibility for the content of the published article.

**Supplementary 2: Detailed Specification of Analytical Dimensions**

**Clinical Function Categories — Definitions and Coding Rules.** The clinical function taxonomy was applied as a primary label to each trial based on a hierarchical reading of the \texttt{interventionsModule}, \texttt{descriptionModule.briefSummary}, and \texttt{outcomesModule.primaryOutcomes} fields. Diagnostic AI was assigned when the primary stated purpose was detection, identification, classification, or staging of a pathological condition from a biological or imaging signal. Predictive and prognostic AI was assigned when the system estimated a future event probability, risk score, or trajectory, even if the prediction was used to guide a downstream clinical decision. Clinical decision support was distinguished from predictive AI by the presence of an explicit therapeutic recommendation, triage output, or management action as the primary deliverable. Generative AI was assigned when the system's output was a natural language artifact --- a note, summary, report, or conversational response --- irrespective of whether the underlying architecture was declared. Agentic AI required evidence of adaptive, closed-loop, or semi-autonomous operation in which the system could modify its own outputs or actions based on evolving patient state without per-step human authorization. Operational AI was assigned to systems whose primary target was healthcare process efficiency rather than patient-level clinical outcomes. Multi-label assignment was permitted where a single trial clearly spanned two functional categories; in practice, fewer than 8% of trials required multi-label coding.

**Data Modality Taxonomy — Full Specifications.** Medical imaging encompassed radiological imaging (plain film, CT, MRI, PET, PET-CT, SPECT), ultrasound modalities (B-mode, Doppler, echocardiography, point-of-care, endoscopic), ophthalmological imaging (fundus photography, OCT, OCT-angiography), dermatological imaging (dermoscopy, reflectance confocal microscopy), endoscopic imaging (colonoscopy, gastroscopy, bronchoscopy, capsule endoscopy, laparoscopic video), pathology imaging (whole slide imaging, digital histopathology, cytology, immunohistochemistry), dental and maxillofacial imaging, and cardiac imaging (cardiac MRI, coronary angiography). Physiological signals encompassed cardiac signals (ECG, Holter, intracardiac electrograms), neurological signals (EEG, ECoG, MEG, evoked potentials), muscular signals (EMG, nerve conduction), respiratory signals (spirometry, polysomnography, capnography), hemodynamic signals (continuous blood pressure, pulse oximetry, PPG), and ocular signals (ERG, EOG). Laboratory and biochemical data included routine laboratory panels, immunological assays, microbiology cultures and PCR, toxicology, point-of-care testing, and biomarker panels (troponins, tumor markers, inflammatory markers, neurodegeneration markers). Omics data encompassed genomics (WGS, WES, SNP arrays, CNV), transcriptomics (bulk and single-cell RNA-seq, spatial transcriptomics), proteomics (mass spectrometry, PEA), metabolomics (NMR- and LC-MS-based), epigenomics (DNA methylation, ATAC-seq, ChIP-seq), microbiomics (16S rRNA, shotgun metagenomics), and multi-omics integrated analyses. Clinical text encompassed structured notes (discharge summaries, operative notes, radiology and pathology reports), unstructured free-text (physician notes, referral letters), patient-generated text (symptom diaries, PRO free text), administrative text (ICD narratives, claims text), and scientific literature. Structured EHR data included coded diagnoses (ICD-10,

SNOMED-CT), medication records, procedure codes, vital signs, demographic and social variables, allergy records, vaccination records, and family history. Wearable and remote monitoring data included consumer wearables (accelerometry, HRV, sleep staging), clinical-grade wearables (ambulatory ECG patches, CGM, implantable loop recorders), implantable device telemetry (pacemakers, ICDs, neurostimulators), smart home sensors, and smartphone-based sensing. Behavioral and cognitive data included neuropsychological assessments, psychiatric rating scales, functional assessments, digital behavioral biomarkers (keyboard dynamics, app usage), eye tracking, and voice and speech features. Patient-reported outcomes included generic quality-of-life instruments (SF-36, EQ-5D, PROMIS), disease-specific PROs, symptom burden scales, treatment satisfaction surveys, and ecological momentary assessment data. Administrative, claims, and registry data included insurance claims, population registries, health system operational data, and socioeconomic and environmental variables. Surgical and procedural data included intraoperative video streams, robotic surgery telemetry, anesthesia monitoring streams (BIS, anesthetic agent concentrations, hemodynamic waveforms), and surgical annotation labels.

**Clinical Specialty Mapping — Domain Definitions and Trial Counts.** Specialty assignment was based on the \texttt{conditionsModule.conditions} and \texttt{conditionsModule.keywords} fields, supplemented by keyword parsing of the brief summary and intervention description where primary specialty was ambiguous. Multi-specialty assignment was permitted and applied to trials operating at specialty interfaces (e.g., cardio-oncology, neuro-oncology). Full trial counts by specialty were: Otolaryngology/ENT 1,994; Oncology 1,295; Cardiology 750; Neurology 637; Endocrinology 579; Pulmonology 534; Gastroenterology 440; Radiology 417; Psychiatry 414; with Rheumatology, Hematology, Infectious Diseases, Medical Genetics, and Allergy/Immunology together contributing fewer than 7% of the total corpus despite representing a substantial collective global burden of disease.

**Appendix D: Integration and Autonomy Level — Taxonomy, Subclassifications, and Coding Procedure.** The four-level taxonomy was applied using a primary coding pass based on the \texttt{interventionsModule} and \texttt{designModule}, followed by a secondary refinement pass using the \texttt{descriptionModule.detailedDescription} and \texttt{outcomesModule.primaryOutcomes} fields. Level 1 required explicit language indicating silent deployment, shadow mode, or observational-only AI operation with no clinical output delivered to care teams. Level 2 assignment required evidence that the AI produced a recommendation, score, alert, or classification delivered to a clinician, who retained authority over the resulting clinical action; sub-classification into A1 versus A2 required determination of whether a mandatory human review step was specified in the protocol. Level 3 required evidence of direct workflow embedding --- automated order entry, alert-triggered pathway activation, or triage re-routing --- without per-encounter clinician override being structurally required. Level 4 required evidence of closed-loop or adaptive operation in which AI system outputs directly modified a physiological intervention parameter (infusion rate, ventilator setting, stimulation amplitude) without mandatory per-adjustment human authorization. The 184 Level 4 trials identified in this corpus span closed-loop insulin delivery ($n \approx 130$), closed-loop anesthesia and sedation control ($n \approx 10$), automated respiratory and oxygen management ($n \approx 10$), closed-loop vasopressor and hemodynamic control ($n \approx 6$), closed-loop antibiotic delivery ($n = 1$, NCT04053140), and closed-loop neurostimulation and brain-computer interfaces ($n \approx 8$). A notable frontier case is NCT07132151, which deploys a reinforcement learning algorithm for fully autonomous ciprofol dosing during surgical abortion without anesthesiologist involvement in individual dosing decisions --- the first prospectively registered RCT to evaluate a self-adapting, non-deterministic learning agent as the autonomous actor in an anesthesia closed-loop system.

**Translational Maturity Stage — Definitions and Boundary Cases.** Retrospective validation was assigned to trials that applied AI to previously collected data without any prospective enrollment or real-time clinical deployment. Silent prospective evaluation was assigned to trials in which the AI system observed real-time clinical data but its outputs were concealed from care teams, enabling performance characterization without influencing care. Interventional decision-support trials required prospective enrollment with the AI system actively providing recommendations to clinicians or patients, with the trial designed to measure the effect of those recommendations on clinical processes or outcomes. Workflow-integrated deployment was assigned to trials in which the AI system was embedded as a routine component of care delivery across an institution or network, with evaluation conducted through pragmatic or stepped-wedge designs. Semi-autonomous clinical operation required closed-loop or adaptive AI operation as defined under Level 4 of the integration and autonomy taxonomy. The boundary between interventional decision-support and workflow-integrated deployment was adjudicated on the basis of whether the trial protocol described the AI as an investigational intervention (former) or as an already-operational system being pragmatically evaluated (latter). The discrepancy between the strict maturity-stage label count of 6 semi-autonomous trials and the autonomy-taxonomy count of 184 Level 4 trials arises from inconsistent free-text labeling in the source data, and the autonomy taxonomy figure should be treated as the more reliable estimate.

**AI Epistemic Role — Tripartite Taxonomy and Regulatory Implications.** The epistemic role dimension was introduced to address a classification gap in existing regulatory and methodological frameworks, which treat AI primarily as a medical device (FDA SaMD, EU AI Act) and therefore focus exclusively on AI-in-the-loop configurations. Level B trials --- in which AI is applied post hoc to trial data --- are methodologically consequential because they determine what inferences can be drawn from the trial corpus, yet they are invisible to device regulators and are rarely distinguished from Level A trials in narrative reviews. Level C trials --- in which AI contributed upstream to the design or discovery of the intervention being evaluated --- are the fastest-growing category in pharmaceutical AI and precision oncology pipelines, yet their AI content is entirely absent from the trial record as conventionally read. The A1/A2 sub-classification within Level A maps directly onto existing regulatory risk tiers: A1 systems correspond to the lower-risk SaMD categories that require demonstration of safety and performance but not pre-market approval in most jurisdictions, while A2 systems correspond to the high-risk autonomous device category that faces the most stringent pre-market scrutiny under both FDA and EU frameworks. Multi-label coding across epistemic levels was applied in 6.2% of trials where a single protocol clearly spanned two roles, most commonly an A1/B hybrid in which the AI tool was the primary intervention and a secondary exploratory aim deployed machine learning to identify responder subgroups.